\documentclass[sigconf,nonacm]{acmart}

\usepackage{booktabs} 

\usepackage{array}
\newcommand{\PreserveBackslash}[1]{\let\temp=\\#1\let\\=\temp}
\newcolumntype{C}[1]{>{\PreserveBackslash\centering}p{#1}}
\newcolumntype{R}[1]{>{\PreserveBackslash\raggedleft}p{#1}}
\newcolumntype{L}[1]{>{\PreserveBackslash\raggedright}p{#1}}

\usepackage{amsmath}
\usepackage{bm}
\usepackage{amsfonts}
\usepackage{amsthm}
\usepackage{multirow}
\usepackage{colortbl}
\usepackage{enumitem}
\usepackage{subcaption}
\usepackage{tabu}
\usepackage{balance}

\def\eqref#1{equation~\ref{#1}}

\def\1{\bm{1}}

\DeclareMathAlphabet{\mathsfit}{\encodingdefault}{\sfdefault}{m}{sl}
\SetMathAlphabet{\mathsfit}{bold}{\encodingdefault}{\sfdefault}{bx}{n}

\DeclareMathOperator*{\bertcat}{BERT_\text{CAT}}
\DeclareMathOperator*{\bertdot}{BERT_\text{DOT}}
\DeclareMathOperator*{\bert}{BERT}
\DeclareMathOperator*{\tff}{TF}
\DeclareMathOperator*{\tk}{TK}
\DeclareMathOperator*{\colbert}{ColBERT}
\DeclareMathOperator*{\prett}{PreTT}

\usepackage{array, booktabs}
\usepackage{graphicx}

\usepackage{pifont}
\usepackage[noabbrev]{cleveref}

\author{Sebastian Hofst{\"a}tter, Sophia Althammer, Michael Schr{\"o}der, Mete Sertkan and Allan Hanbury}
\email{first.last@tuwien.ac.at}
\affiliation{\institution{TU Wien, Vienna, Austria}}

\begin{document}

\title{Improving Efficient Neural Ranking Models with Cross-Architecture Knowledge Distillation}

\begin{abstract}

Retrieval and ranking models are the backbone of many applications such as web search, open domain QA, or text-based recommender systems. The latency of neural ranking models at query time is largely dependent on the architecture and deliberate choices by their designers to trade-off effectiveness for higher efficiency. This focus on low query latency of a rising number of efficient ranking architectures make them feasible for production deployment. In machine learning an increasingly common approach to close the effectiveness gap of more efficient models is to apply knowledge distillation from a large teacher model to a smaller student model. We find that different ranking architectures tend to produce output scores in different magnitudes. Based on this finding, we propose a cross-architecture training procedure with a margin focused loss (Margin-MSE), that adapts knowledge distillation to the varying score output distributions of different BERT and non-BERT passage ranking architectures. We apply the teachable information as additional fine-grained labels to existing training triples of the MSMARCO-Passage collection. We evaluate our procedure of distilling knowledge from state-of-the-art concatenated BERT models to four different efficient architectures (TK, ColBERT, PreTT, and a BERT CLS dot product model). We show that across our evaluated architectures our Margin-MSE knowledge distillation significantly improves re-ranking effectiveness without compromising their efficiency. Additionally, we show our general distillation method to improve nearest neighbor based index retrieval with the BERT dot product model, offering competitive results with specialized and much more costly training methods. To benefit the community, we publish the teacher-score training files in a ready-to-use package.

\end{abstract}

\maketitle

\section{Introduction}

The same principles that applied to traditional IR systems to achieve low query latency also apply to novel neural ranking models: We need to transfer as much computation and data transformation to the indexing phase as possible to require less resources at query time \cite{mackenzie2020efficiency,manning2008introduction}. For the most effective BERT-based \cite{devlin2018bert} neural ranking models, which we refer to as \textbf{BERT$_\textbf{CAT}$}, this transfer is simply not possible, as the concatenation of query and passage require all Transformer layers to be evaluated at query time to receive a ranking score \cite{nogueira2019passage}. 

To overcome this architecture restriction the neural-IR community proposed new architectures by deliberately choosing to trade-off effectiveness for higher efficiency. Among these low query latency approaches are: \textbf{TK} \cite{Hofstaetter2020_ecai} with shallow Transformers and separate query and document contextualization; \textbf{ColBERT}~\cite{khattab2020colbert} with late-interactions of BERT term representations; \textbf{PreTT}~\cite{macavaney2020efficient} with a combination of query-independent and query-dependent Transformer layers; and a BERT-CLS dot product scoring model which we refer to as \textbf{BERT$_\textbf{DOT}$}, also known in the literature as Tower-BERT \cite{chang2020pre}, BERT-Siamese \cite{xiong2020approximate}, or TwinBERT \cite{lu2020twinbert}.\footnote{Yes, we see the irony: \url{https://xkcd.com/927/}} Each approach has unique characteristics that make them suitable for production-level query latency which we discuss in Section \ref{sec:models}. 

An increasingly common way to improve smaller or more efficient models is to train them, as students, to imitate the behavior of larger or ensemble teacher models via Knowledge Distillation (KD) \cite{hinton2015distilling}. This is typically applied to the same architecture with fewer layers and dimensions \cite{jiao2019tinybert,sanh2019distilbert} via the output or layer-wise activations \cite{sun2019patient}. KD has been applied in the ranking task for the same architecture with fewer layers \cite{li2020parade,gao2020understanding,chen2020simplified} and in constrained sub-tasks, such as keyword-list matching \cite{lu2020twinbert}.

In this work we propose a model-agnostic training procedure using cross-architecture knowledge distillation from $\bertcat$ with the goal to improve the effectiveness of efficient passage ranking models without compromising their query latency benefits. 

\begin{figure}[t]
   \includegraphics[trim={0.4cm 0cm 0cm 0cm},clip,width=0.47\textwidth]{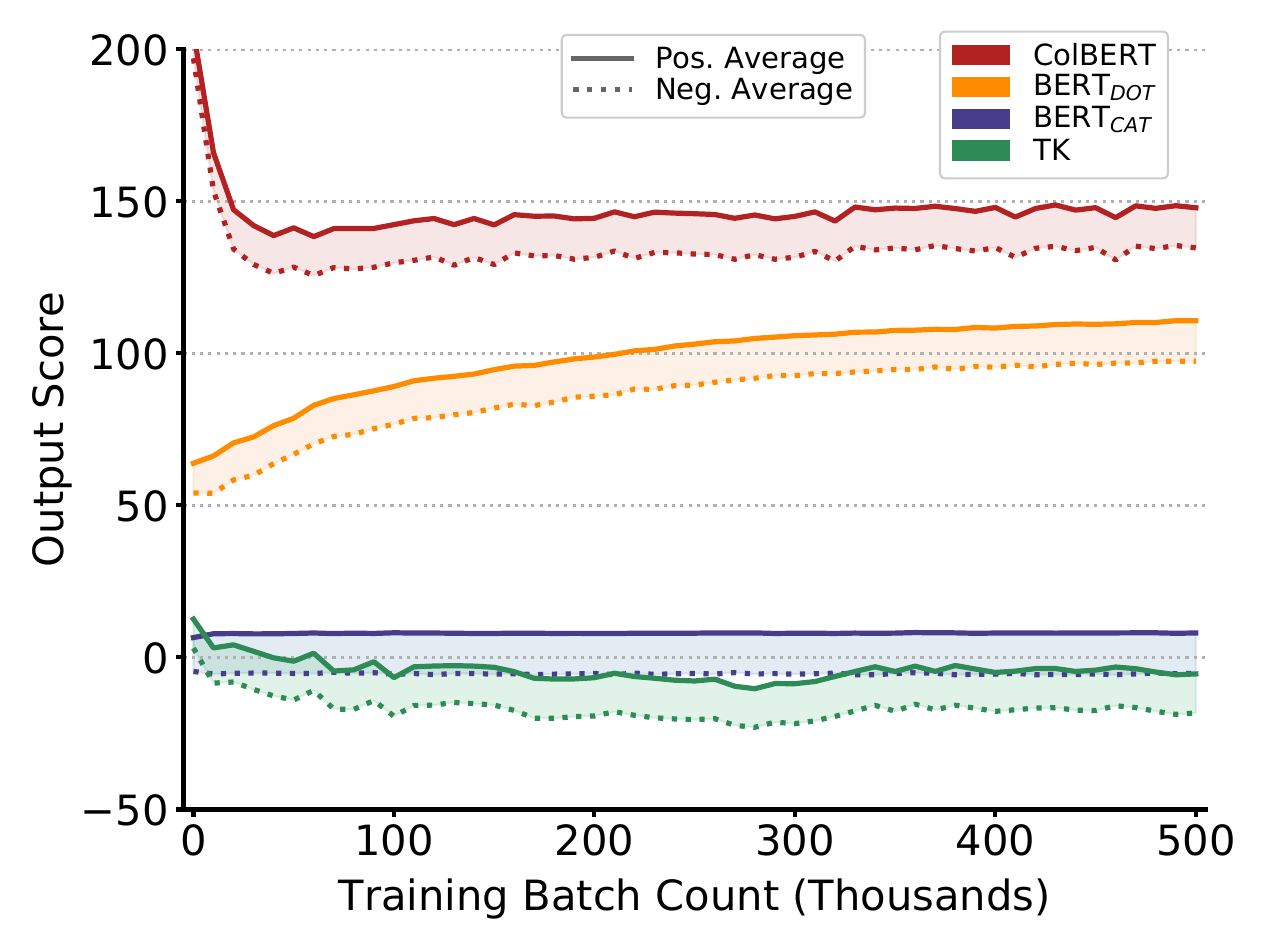}
    \centering
    \vspace{-0.4cm}
    \caption{Raw query-passage pair scores during training of different ranking models. The margin between the positive and negative samples is shaded.}
    \label{fig:student_pos_neg_score_training}
    \vspace{-0.5cm}
\end{figure}

A unique challenge for knowledge distillation in the ranking task is the possible range of scores, i.e. a ranking model outputs a single unbounded decimal value and the final result solely depends on the relative ordering of the scores for the candidate documents per query. We make the crucial observation, depicted in Figure \ref{fig:student_pos_neg_score_training}, that different architectures during their training gravitate towards unique range patterns in their output scores. The $\bertcat$ model exhibits positive relevant-document scores, whereas on average the non-relevant documents are below zero. The $\tk$ model solely produces negative averages, and the $\bertdot$ and $\colbert$ models, due to their dot product scoring, show high output scores. This leads us to our main research question:

\newcommand{\RQone}{\begin{itemize}
    \item[\textbf{RQ1}] How can we apply knowledge distillation in retrieval across architecture types?
\end{itemize}}
\newcommand{\RQoneRunning}{\textbf{RQ1} \textit{How can we apply knowledge distillation in retrieval across architecture types? }}
\RQone

To optimally support the training of cross-architecture knowledge distillation, we allow our models to converge to a free scoring range, as long as the margin is alike with the teacher. We make use of the common triple \textit{(q, relevant doc, non-relevant doc)} training regime, by distilling knowledge via the margin of the two scoring pairs. We train the students to learn the same margin as their teachers, which leaves the models to find the most \textit{comfortable} or natural range for their architecture. We optimize the student margin to the teacher margin with a Mean Squared Error loss (Margin-MSE). We confirm our strategy with an ablation study of different knowledge distillation losses and show the Margin-MSE loss to be the most effective.

Thanks to the rapid advancements and openness of the Natural Language Processing community, we have a number of pre-trained BERT-style language models to choose from to create different variants of the $\bertcat$ architecture to study, allowing us to answer:

\newcommand{\RQtwo}{\begin{itemize}
    \item[\textbf{RQ2}] How effective is the distillation with a single teacher model in comparison to an ensemble of teachers?
\end{itemize}}
\RQtwo

We train three different $\bertcat$ versions as teacher models with different initializations: BERT-Base \cite{devlin2018bert}, BERT-Large with whole word masking \cite{devlin2018bert}, and ALBERT-large \cite{lan2019albert}.
To understand the behavior that the different language models bring to the $\bertcat$ architecture, we compare their training score margin distributions and find that the models offer variability suited for an ensemble. 

We created the teacher ensemble by averaging each of the three scores per query-document pair. We conduct the knowledge distillation with a single teacher and a teacher ensemble. The knowledge distillation has a general positive effect on all retrieval effectiveness metrics of our student models. In most cases the teacher ensemble further improves the student models' effectiveness in the re-ranking scenario above the already improved single teacher training.

The dual-encoder $\bertdot$ model can be used for full collection indexing and retrieval with a nearest neighbor vector search approach, so we study:

\newcommand{\RQtwoB}{\begin{itemize}
    \item[\textbf{RQ3}] How effective is our distillation for dense nearest neighbor retrieval?
\end{itemize}}
\RQtwoB

We observe similar trends in terms of effectiveness per teacher strategy, with increased effectiveness of $\bertdot$ models for a single teacher and again a higher increase for the ensemble of teachers. Even though we do not add dense retrieval specific training methods, such as index-based passage sampling \cite{xiong2020approximate} or in-batch negatives \cite{lin2020distilling} we observe very competitive results compared to those much more costly training approaches. 

To put the improved models in the perspective of the efficiency-effectiveness trade-off, we investigated the following question:

\newcommand{\RQthree}{\begin{itemize}
    \item[\textbf{RQ4}] By how much does effective knowledge distillation shift the balance in the efficiency-effectiveness trade-off?
\end{itemize}}
\RQthree

We show how the knowledge distilled efficient architectures outperform the $\bertcat$ baselines on several metrics. There is no longer a compromise in utilizing $\prett$ or $\colbert$ and the effectiveness gap, i.e. the difference between the most effective and the other models, of $\bertdot$ and $\tk$ is significantly smaller.

The contributions of this work are as follows:
\begin{itemize}
    \item We propose a cross-architecture knowledge distillation procedure with a Margin-MSE loss for a range of neural retrieval architectures
    \item We conduct a comprehensive study of the effects of cross-architecture knowledge distillation in the ranking scenario
    \item We publish our source code as well as ready-to-use teacher training files for the community at: \\ \url{https://github.com/sebastian-hofstaetter/neural-ranking-kd} %
\end{itemize}

\section{Retrieval Models}
\label{sec:models}

\begin{table*}[t!]
    \centering
    \caption{Comparison of model characteristics using DistilBERT instances. \textit{Effectiveness} compares the baseline nDCG@10 of MSMARCO-DEV. \textit{NN Index} refers to indexing the passage representations in a nearest neighbor index. $|P|$ refers to the number of passages; $|T|$ to the total number of term occurrences in the collection; $m$ the query length; and $n$ the document length.}
    \label{tab:model_overview}
    \vspace{-0.3cm}
    \setlength\tabcolsep{5.8pt}
    \begin{tabular}{llrr!{\color{lightgray}\vrule}llll}
       \toprule
       \multirow{2}{*}{\textbf{Model}} & \multirow{2}{*}{\textbf{Effectiveness}} & \textbf{Query} & \textbf{GPU} & \multirow{2}{*}{\textbf{Query-Passage Interaction}} & \textbf{Passage} & \textbf{NN} & \textbf{Storage Req. }  \\
       &&\textbf{Latency} & \textbf{Memory} & & \textbf{Cache} & \textbf{Index}& \textbf{($\times$ Vector Size)} \\
        \midrule
        \textbf{BERT$_\text{CAT}$} & 1  & 950 ms & 10.4 GB & All TF layers           & --          & --          & --            \\
        \midrule
        \textbf{BERT$_\text{DOT}$} & $\times$ 0.87 & 23 ms & 3.6 GB & Single dot product              & \checkmark & \checkmark & $|P|$       \\ 
        \textbf{ColBERT}           & $\times$ 0.97 & 28 ms & 3.4 GB & $m * n$ dot products    & \checkmark & \checkmark & $|T|$  \\
        \textbf{PreTT}             & $\times$ 0.97 & 455 ms & 10.9 GB & Min. 1 TF layer (here 3)                 & \checkmark & -- & $|T|$  \\
        \textbf{TK}                & $\times$ 0.89 & 14 ms & 1.8 GB & $m * n$ dot products + Kernel-pooling    & \checkmark & -- & $|T|$  \\
        \bottomrule
    \end{tabular}
    \vspace{-0.2cm}
\end{table*}

We study the effects of knowledge distillation on a wide range of recently introduced Transformer- \& BERT-based ranking models. We describe their architectures in detail below and summarize them in Table \ref{tab:model_overview}.

\subsection{BERT\texorpdfstring{$_\textbf{CAT} $} : Concatenated Scoring}

The common way of utilizing the BERT pre-trained Transformer model in a re-ranking scenario \cite{nogueira2019passage,macavaney2019,yilmaz2019cross} is by concatenating query and passage input sequences. We refer to this base architecture as BERT$_\text{CAT}$. In the BERT$_\text{CAT}$ ranking model, the query ${q}_{1:m}$ and passage ${p}_{1:n}$ sequences are concatenated with special tokens (using the $;$ operator) and the CLS token representation computed by BERT (selected with $_1$) is scored with single linear layer $W_s$:
\begin{equation}
\begin{aligned}
    \bertcat({q}_{1:m},{p}_{1:n}) & = \bert([\text{CLS};{q}_{1:m};\text{SEP};{p}_{1:n}])_1 * W_s
\end{aligned}
\end{equation}

We utilize BERT$_\text{CAT}$ as our teacher architecture, as it represents the current state-of-the art in terms of effectiveness, however it requires substantial compute at query time and increases the query latency by seconds \cite{Hofstaetter2019_osirrc,xiong2020approximate}. Simply using smaller BERT variants does not change the design flaw of having to compute every representation at query time.

\subsection{BERT\texorpdfstring{$_\textbf{DOT} $} : Dot Product Scoring}

In contrast to BERT$_\text{CAT}$, which requires a full online computation, the BERT$_\text{DOT}$ model only matches a single CLS vector of the query with a single CLS vector of a passage \cite{xiong2020approximate,luan2020sparse,lu2020twinbert}. 
The BERT$_\text{DOT}$ model uses two independent $\bert$ computations as follows:
\begin{equation}
\begin{aligned} 
\hat{q} &= \bert([\text{CLS};{q}_{1:m}])_1 * W_s  \\
\hat{p} &= \bert([\text{CLS};{p}_{1:n}])_1 * W_s
\end{aligned}
\end{equation}
which allows us to pre-compute every contextualized passage representation $\hat{p}$. After this, the model computes the final scores as the dot product $\cdot$ of $\hat{q}$ and $\hat{p}$:
\begin{equation}
\begin{aligned}
    \bertdot({q}_{1:m},{p}_{1:n}) & = \hat{q} \cdot \hat{p}
\end{aligned}
\end{equation}

BERT$_\text{DOT}$, with its bottleneck of comparing single vectors, compresses information much more strongly than BERT$_\text{CAT}$, which brings large query time improvements at the cost of lower effectiveness, as can be seen in Table \ref{tab:model_overview}. 

\subsection{ColBERT}

The $\colbert$ model \cite{khattab2020colbert} is similar in nature to $\bertdot$, by delaying the interactions between query and document to after the BERT computation. $\colbert$ uses every query and document representation:
\begin{equation}
\begin{aligned} 
\hat{q}_{1:m} &= \bert([\text{CLS};{q}_{1:m};\operatorname{rep}(\text{MASK})]) * W_s  \\
\hat{p}_{1:n} &= \bert([\text{CLS};{p}_{1:n}]) * W_s
\end{aligned}
\end{equation}
where the $\operatorname{rep}(MASK)$ method repeats the MASK token a number of times, set by a hyperparameter. \citet{khattab2020colbert} introduced this query augmentation method to increase the computational capacity of the BERT model for short queries. We independently confirmed that adding these MASK tokens improves the effectiveness of $\colbert$. 
The interactions in the $\colbert$ model are aggregated with a max-pooling per query term and sum of query-term scores as follows:
\begin{equation}
\begin{aligned}
    \colbert({q}_{1:m},{p}_{1:n}) = \sum_{1}^{m} \max_{1..n} \hat{q}_{1:m}^T \cdot \hat{p}_{1:n}
\end{aligned}
\end{equation}
The aggregation only requires $n*m$ dot product computations, making it roughly as efficient as $\bertdot$, however the storage cost of pre-computing passage representations is much higher and depends on the total number of terms in the collection. \citet{khattab2020colbert} proposed to compress the dimensions of the representation vectors by reducing the output features of $W_s$. We omitted this compression, as storage space is not the focus of our study and to better compare results across different models.

\subsection{PreTT}

The $\prett$ architecture \cite{macavaney2020efficient} is conceptually between $\bertcat$ and $\colbert$, as it allows to compute $b$ BERT-layers separately for query and passage:

\begin{equation}
\begin{aligned} 
\hat{q}_{1:m} &= \bert_{1:b}([\text{CLS};{q}_{1:m}])  \\
\hat{p}_{1:n} &= \bert_{1:b}([\text{CLS};{p}_{1:n}])
\end{aligned}
\end{equation}

Then $\prett$ concatenates the sequences with a SEP separator token and computes the remaining layers to compute a total of $\hat{b}$ BERT-layers. Finally, the CLS token output is pooled with single linear layer $W_s$:

\begin{equation}
\begin{aligned}
    \prett({q}_{1:m},{p}_{1:n}) & = \bert_{b:\hat{b}}([\hat{q}_{1:m};\text{SEP};\hat{p}_{1:n}])_1 * W_s
\end{aligned}
\end{equation}

Concurrently to $\prett$, DC-BERT \cite{zhang2020dc} and EARL \cite{gao2020earl} have been proposed with very similar approaches to split Transformer layers. We selected $\prett$ simply as a representative of this group of models. Similar to $\colbert$, we omitted the optional compression of representations for better comparability.  

\subsection{Transformer-Kernel}

The Transformer-Kernel (TK) model \cite{Hofstaetter2020_ecai} is not based on BERT pre-training, but rather uses shallow Transformers. TK independently contextualizes query ${q}_{1:m}$ and passage ${p}_{1:n}$ based on pre-trained word embeddings, where the intensity of the contextualization (Transformers as $\tff$) is set by a gate $\alpha$:
\begin{equation}
\begin{aligned} 
\hat{q}_i &= q_i * \alpha + \tff(q_{1:m})_i * (1 - \alpha) \\
\hat{p}_i &= p_i * \alpha + \tff(p_{1:n})_i * (1 - \alpha)
\label{eq:representation}
\end{aligned}
\end{equation}

The sequences $\hat{q}_{1:m}$ and $\hat{p}_{1:n}$ interact in a match-matrix with a cosine similarity per term pair and each similarity is activated by a set of Gaussian kernels \cite{Xiong2017}:
\begin{equation}
K^{k}_{i,j} = \exp \left(-\frac{\left(\cos(\hat{q_i},\hat{p_j})-\mu_{k}\right)^{2}}{2 \sigma^{2}}\right)
\end{equation}
Kernel-pooling is a soft-histogram, which counts the number of occurrences of similarity ranges. Each kernel $k$ focuses on a fixed range with center $\mu_{k}$ and width of $\sigma$.

These kernel activations are then summed, first by the passage term dimension $j$, log-activated, and then the query dimension is summed, resulting in a single score per kernel. The final score is calculated by a weighted sum using $W_s$:
\begin{equation}\label{eq:rsv}
\tk({q}_{1:m},{p}_{1:n}) = \bigg(\sum_{i=1}^{m} \log\left( \sum_{j=1}^{n} K^{k}_{i,j} \right) \bigg) * W_s
\end{equation}

\begin{figure*}[t]
   \includegraphics[trim={0.3cm 0.3cm 0.3cm 0.7cm},clip,width=0.8\textwidth]{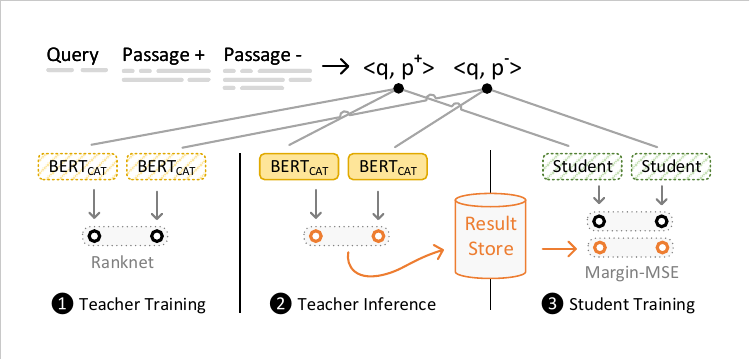}
    \centering
    \vspace{-0.6cm}
    \caption{Our knowledge distillation process, re-visiting the same training triples in all steps: \ding{202} Training the \textbf{BERT$_\textbf{CAT}$} model; \ding{203} Using the trained \textbf{BERT$_\textbf{CAT}$} to create scores for all training triples; \ding{204} Individually training the student models with Margin-MSE using the teacher scores.}
    \label{fig:kd_process}
   \vspace{-0.3cm}
\end{figure*}

\subsection{Comparison}

In Table \ref{tab:model_overview} we summarize our evaluated models. We compare the efficiency and effectiveness trade-off in the leftmost section, followed by a general overview of the model capabilities in the right most section. We measure the query latency for 1 query and 1000 documents with cached document representations where applicable and report the peak GPU memory requirement for the inference of the validation set. We summarize our observations of the different model characteristics:
\begin{itemize}
    \item The query latency of $\bertcat$ is prohibitive for efficient production use (Except for head queries that can be fully pre-computed).
    \item $\bertdot$ is the most efficient BERT-based model with regards to storage and query latency, at the cost of lower effectiveness compared to ColBERT and PreTT.
    \item PreTT highly depends on the choice of the concatenation-layer hyperparameter, which we set to 3 to be between $\bertcat$ and $\colbert$.
    \item $\colbert$ is especially suited for small collections, as it requires a large passage cache.
    \item $\tk$ is less effective overall, however it is much cheaper to run than the other models.
\end{itemize}

The most suitable neural ranking model ultimately depends on the exact scenario.
To allow people to make the choice, we evaluated all presented models. we use $\bertcat$ as our teacher architecture and the other presented architectures as students.

\section{Cross-Architecture Knowledge~Distillation}
\label{sec:cross-arch}

The established approach to training deep neural ranking models is mainly based on large-scale annotated data. Here, the MSMARCO collection is becoming the de-facto standard. The MSMARCO collection only contains binary annotations for fewer than two positive examples per query, and no explicit annotations for non-relevant passages. The approach proposed by \citet{msmarco16} is to utilize randomly selected passages retrieved from the top 1000 candidates of a traditional retrieval system as negative examples. This approach works reasonably well, but accidentally picking relevant passages is possible. 

Neural retrieval models are commonly trained on triples of binary relevance assignments of one relevant and one non-relevant passage. However, they are used in a setting that requires a much more nuanced view of relevance when they re-rank a thousand possibly relevant passages. The $\bertcat$ architecture shows the strongest generalization capabilities, which other architectures do not posses. 

Following our observation of distinct scoring ranges of different model architectures in Figure \ref{fig:student_pos_neg_score_training}, we propose to utilize a knowledge distillation loss by only optimizing the margin between the scores of the relevant and the non-relevant sample passage per query. We call our proposed approach Margin Mean Squared Error (Margin-MSE). We train ranking models on batches containing triples of queries $Q$, relevant passages $P^+$, and non-relevant passages $P^-$. We utilize the output margin of the teacher model $M_t$ as label to optimize the weights of the student model $M_s$: 

\begin{equation}
\begin{aligned} 
\mathcal{L}(Q,P^{+},P^{-}) = \operatorname{MSE}(&M_s(Q,P^{+}) - M_s(Q,P^{-}),\\ &M_t(Q,P^{+}) - M_t(Q,P^{-}))
\end{aligned}
\label{eq:loss}
\end{equation}

MSE is the Mean Squared Error loss function, calculating the mean of the squared differences between the scores $S$ and the targets $T$ over the batch size:

\begin{equation}
\operatorname{MSE}(S,T) = \frac{1}{|S|} \sum_{s \in S, t \in T} (s - t)^2
\label{eq:mse}
\end{equation}

The Margin-MSE loss discards the original binary relevance information, in contrast to other knowledge distillation approaches \cite{li2020parade}, as the margin of the teacher can potentially be negative, which would indicate a reverse ordering from the original training data. We observe that the teacher models have a very high pairwise ranking accuracy during training of over 98\%, therefore we view it as redundant to add the binary information in the ranking loss.\footnote{We do not analyze this statistic further in this paper, as we did not see a correlation or interesting difference between models on this pairwise training accuracy metric.}

In Figure \ref{fig:kd_process} we show the staged process of our knowledge distillation. For simplicity and ease of re-use, we utilize the same training triples for every step. The process begins with training a $\bertcat$ teacher model on the collection labels with a RankNet loss \cite{burges2010ranknet}. After the teacher training is finished, we use the teacher model again to infer all scores for the training data, without updating its weights. This allows us to store the teacher scores once, for an efficient experimentation and sharing workflow. Finally, we train our student model of a different architecture, by using the teacher scores as labels with our proposed Margin-MSE loss.

\section{Experiment design}

For our neural re-ranking training and inference we use PyTorch~\cite{pytorch2017} and the HuggingFace Transformer library \cite{wolf2019huggingface}. For the first stage indexing and retrieval we use Anserini \cite{Yang2017}.

\subsection{Collection \& Query Sets}

We use the MSMARCO-Passage~\cite{msmarco16} collection with sparsely-judged MSMARCO-DEV query set of 49,000 queries as well as the densely-judged query set of 43 queries derived from TREC-DL'19 \cite{trec2019overview}. For TREC graded relevance labels we use a binarization point of 2 for MRR and MAP. MSMARCO is based on sampled Bing queries and contains 8.8 million passages with a proposed training set of 40 million triples sampled. We evaluate our teachers on the full training set, so to not limit future work in terms of the number of triples available. We cap the query length at $30$ tokens and the passage length at $200$ tokens.

\subsection{Training Configuration}

We use the Adam \cite{kingma2014adam} optimizer with a learning rate of $7*10^{-6}$ for all BERT layers, regardless of the number of layers trained. TK is the only model trained on a higher rate of $10^{-5}$. We employ early stopping, based on the best nDCG@10 value of the validation set. We use a training batch size of 32.

\subsection{Model Parameters}

All student language models use a 6-layer DistilBERT \cite{sanh2019distilbert} as their initialization standpoint. We chose DistilBERT over BERT-Base, as it has been shown to provide a close lower bound on the results at half the runtime \cite{sanh2019distilbert,macavaney2020efficient}. For our ColBERT implementation we repeat the query MASK augmentation 8 times, regardless of the amount of padding in a batch in contrast to \citet{khattab2020colbert}. For PreTT we decided to concatenate sequences after 3 layers of the 6 layer DistilBERT, as we want to evaluate it as a mid-choice between ColBERT and $\bertcat$. For TK we use the standard 2 layer configuration with 300 dimensional embeddings. For the traditional BM25 we use the tuned parameters from the Anserini documentation.

\section{Results}

We now discuss our research questions, starting with the study of our proposed Margin-MSE loss function; followed by an analysis of different teacher model results and their impact on the knowledge distillation; and finally examining what the knowledge distillation improvement means for the efficiency-effectiveness trade-off.

\begin{table}[t]
    \centering
    \caption{Loss function ablation results on MSMARCO-DEV, using a single teacher (\textit{T1} in Table \ref{tab:all_results}). The original training baseline is indicated by --.}
    \label{tab:loss_abl}
    \setlength\tabcolsep{2pt}
    \vspace{-0.3cm}
    \begin{tabular}{llrrr}
       
       \toprule
       \textbf{Model} & \textbf{KD Loss} &  nDCG@10 & MRR@10 & MAP@100  \\
       \midrule
       \multirow{4}{*}{ColBERT} & --    & .417 & .357 & .361 \\
                                & Weighted RankNet  & .417 & .356 & .360 \\
                                & Pointwise MSE     & .428 & .365 & .369 \\
                                & Margin-MSE        & \textbf{.431} & \textbf{.370} & \textbf{.374} \\

       \midrule
       \multirow{4}{*}{BERT$_\text{DOT}$} & --     & .373 & .316 & .321 \\
                                                & Weighted RankNet   & .384 & .326 & .332 \\
                                                & Pointwise MSE      & .387 & .328 & .332 \\
                                                & Margin-MSE         & \textbf{.388} & \textbf{.330} & \textbf{.335} \\

       \midrule
       \multirow{4}{*}{TK} & --    & .384 & .326 & .331 \\
                           & Weighted RankNet  & .387 & .328 & .333 \\
                           & Pointwise MSE     & .394 & .335 & .340 \\
                           & Margin-MSE        & \textbf{.398} & \textbf{.339} & \textbf{.344} \\

       \bottomrule
    \vspace{-.6cm}
    \end{tabular}
\end{table}

\subsection{Optimization Study}

We validate our approach presented in Section \ref{sec:cross-arch} and our research question \RQoneRunning 
by comparing Margin-MSE with different knowledge distillation losses using the same training data. We compare our approach with a pointwise MSE loss, defined as follows:

\begin{equation}
\begin{aligned} 
\mathcal{L}(Q,P^{+},P^{-}) = &\operatorname{MSE}(M_s(Q,P^{+}), M_t(Q,P^{+}))\; + \\ &\operatorname{MSE}(M_s(Q,P^{-}), M_t(Q,P^{-}))
\end{aligned}
\label{eq:loss}
\end{equation}

This is a standard approach already used by \citet{vakili2020distilling} and \citet{li2020parade}. Additionally, we utilize a weighted RankNet loss, where we weight the samples in a batch according to the teacher margin:

\begin{equation}
\begin{aligned} 
\mathcal{L}(Q,P^{+},P^{-}) = \operatorname{RankNet}(&M_s(Q,P^{+}) - M_s(Q,P^{-}))\; * \\ ||&M_t(Q,P^{+}) - M_t(Q,P^{-})||
\end{aligned}
\label{eq:loss}
\end{equation}

We show the results of our ablation study in Table \ref{tab:loss_abl} for three distinct ranking architectures that significantly differ from the $\bertcat$ teacher model. We use a single (BERT-Base$_{CAT}$) teacher model for this study. For each of the three architectures the Margin-MSE loss outperforms the pointwise MSE and weighted RankNet losses on all metrics. However, we also note that applying knowledge distillation in general improves each model's result over the respective original baseline. Our aim in proposing to use the Margin-MSE loss was to create a simple yet effective solution that does not require changes to the model architectures or major adaptions to the training procedure.

\begin{table*}[t!]
    \centering
    \caption{Effectiveness results for both query sets of our baselines (results copied from cited models), teacher model results (with the teacher signs left of the model name), and  using those teachers for our student models.}
    \label{tab:all_results}
    \setlength\tabcolsep{4pt}
    \begin{tabular}{cll!{\color{lightgray}\vrule}rrr!{\color{lightgray}\vrule}rrr}
       \toprule
       \multirow{2}{*}{} & \multirow{2}{*}{\textbf{Model}}&
       \multirow{2}{*}{\textbf{Teacher}}   &
       \multicolumn{3}{c!{\color{lightgray}\vrule}}{\textbf{TREC DL Passages 2019}}&
       \multicolumn{3}{c}{\textbf{MSMARCO DEV}}\\
       &&& nDCG@10 & MRR@10 & MAP@1000 & nDCG@10 & MRR@10 & MAP@1000 \\
        \midrule
        \multicolumn{6}{l}{\textbf{Baselines}} \\
         & BM25     & -- &   .501 & .689 & .295 & .241 & .194 & .202 \\
         & TREC Best Re-rank \cite{yan2020idst}     & -- &   .738 & .882 & .457 & -- & -- & -- \\
         & BERT$_\text{CAT}$ (6-Layer Distilled Best) \cite{gao2020understanding}     & -- &   .719 & -- & -- & -- & .356 & -- \\

         & BERT-Base$_\text{DOT}$ ANCE \cite{xiong2020approximate}     & -- &   .677 & -- & -- & -- & .330 & -- \\
        \midrule
        \multicolumn{6}{l}{\textbf{Teacher Models}} \\
        $T1$ & BERT-Base$_\text{CAT}$    & -- &   .730 & .866 & .455 & .437 & .376 & .381 \\
        & BERT-Large-WM$_\text{CAT}$     & -- &   .742 & .860 & .484 & .442 & .381 & .385 \\
        & ALBERT-Large$_\text{CAT}$      & -- &   .738 & \textbf{.903} & .477 & .446 & .385 & .388 \\
        $T2$ & Top-3 Ensemble            & -- &   \textbf{.743} & .889 & \textbf{.495} & \textbf{.460} & \textbf{.399} & \textbf{.402} \\
        
         \midrule
         \multicolumn{6}{l}{\textbf{Student Models}} \\
         &\multirow{3}{*}{DistilBERT$_\text{CAT}$}  & --  & .723 & .851 & .454 & .431 & .372 & .375 \\
         &                                          & T1 & .739 & .889 & .473 & .440 & .380 & .383 \\
         &                                          & T2 & \textbf{.747} & \textbf{.891} & \textbf{.480} &\textbf{ .451} & \textbf{.391} & \textbf{.394} \\

         \arrayrulecolor{lightgray}
         \midrule
         &\multirow{3}{*}{PreTT}                    & -- &   .717 & .862 & .438 & .418 & .358 & .362 \\
         &                                          & T1 &  \textbf{.748} & \textbf{.890} & \textbf{.475} & .439 & .378 & .382 \\
         &                                          & T2 & .737 & .859 & .472 & \textbf{ .447} & \textbf{.386} & \textbf{.389} \\
         
         \midrule
         &\multirow{3}{*}{ColBERT}                  & -- &   .722 & .874 & .445 & .417 & .357 & .361 \\
         &                                          & T1 &  .738 & .862 & .472 & .431 & .370 & .374 \\
         &                                          & T2 &  \textbf{.744} & \textbf{.878} & \textbf{.478} & \textbf{.436} & \textbf{.375} & \textbf{.379} \\

         \midrule
         &\multirow{3}{*}{BERT-Base$_\text{DOT}$}   & -- &   .675 & .825 & .396 & .376 & .320 & .325 \\
         &                                          & T1 &  .677 & .809 & .427 & .378 & .321 & .327 \\
         &                                          & T2 &  \textbf{.724} & \textbf{.876} & \textbf{.448} & \textbf{.390} & \textbf{.333} & \textbf{.338} \\

         \midrule 
         &\multirow{3}{*}{DistilBERT$_\text{DOT}$}  & -- &  .670 & .841 & .406 & .373 & .316 & .321 \\
         &                                          & T1 & .704 & .821 & .441 & .388 & .330 & .335 \\
         &                                          & T2 & \textbf{.712} & \textbf{.862} & \textbf{.453} & \textbf{.391} & \textbf{.332} & \textbf{.337} \\
         \midrule

         &\multirow{3}{*}{TK}                       & -- &   .652 & .751 & .403 & .384 & .326 & .331 \\
         &                                          & T1 &  \textbf{.669} & \textbf{.813} & .414 & .398 & .339 & .344 \\
         &                                          & T2 & .666 & .797 & \textbf{.415} & \textbf{.399} & \textbf{.341} & \textbf{.345} \\

        \arrayrulecolor{black}
        \bottomrule
    \end{tabular}
\end{table*}




\subsection{Knowledge Distillation Results}

Utilizing our proposed Margin-MSE loss in connection with our trained teacher models, we follow the procedure laid out in Section \ref{sec:cross-arch} to train our knowledge-distilled student models. Table \ref{tab:all_results} first shows our baselines, then in the second section the results of our teacher models, and in the third section our student architectures. Each student has a baseline result without teacher training (depicted by --) and a single teacher T1 as well as the teacher ensemble denoted with T2. With these results we can now answer:

\RQtwo

We selected BERT-Base$_\text{CAT}$ as our single teacher model, as it is a commonly used instance in neural ranking models. The ensemble of different larger $\bertcat$ models shows strong and consistent improvements on all MSMARCO DEV metrics and MAP@1000 of TREC-DL'19. When we compare our teacher model results with the best re-ranking entry \cite{yan2020idst} of TREC-DL'19, we see that our teachers, especially the ensemble outperform the TREC results to represent state-of-the-art results in terms of effectiveness. 

Overall, we observe that either a single teacher or an ensemble of teachers improves the model results over their respective original baselines. The ensemble T2 improves over T1 for all models on the sparse MSMARCO-DEV labels with many queries. Only on the TREC-DL'19 query set does T2 fail to improve over T1 for TK and PreTT. The only outlier in our results is BERT-Base$_\text{DOT}$ trained on T1, where there is no improvement over the baseline, T2 however does show a substantial improvement. This leads us to the conclusion that utilizing an ensemble of teachers is overall preferred to a single teacher model. 

Furthermore, when we compare the BERT type for the $\bertcat$ architecture, we see that DistilBERT$_\text{CAT}$-T2 outperforms any single teacher model with twice and four times the layers on almost all metrics. For the $\bertdot$ architecture we also compared BERT-Base and DistilBERT, both as students, and here BERT-Base has a slight advantage trained on T2. However, its T1 results are inconsistent, where almost no improvement is observable, whereas DistilBERT$_\text{DOT}$ exhibits consistent gains first for T1 and then another step for T2.

Our T2 training improves both instances of the $\bertdot$ architecture in comparison to the ANCE \cite{xiong2020approximate} trained $\bertdot$ model and evaluated in the re-ranking setting. 

\begin{table*}[t!]
    \centering
    \caption{Dense retrieval results for both query sets, using a flat Faiss index without compression.}
    \label{tab:dense_results}
    \vspace{-0.3cm}
    \setlength\tabcolsep{4pt}
    \begin{tabular}{clll!{\color{lightgray}\vrule}rrr!{\color{lightgray}\vrule}rrr}
       \toprule
       \multirow{2}{*}{} & \multirow{2}{*}{\textbf{Model}} & \textbf{Index}&
       \multirow{2}{*}{\textbf{Teacher}}   &
       \multicolumn{3}{c!{\color{lightgray}\vrule}}{\textbf{TREC DL Passages 2019}}&
       \multicolumn{3}{c}{\textbf{MSMARCO DEV}}\\
       &&\textbf{Size}&& nDCG@10 & MRR@10 & Recall@1K & nDCG@10 & MRR@10 & Recall@1K \\
        \midrule
        \multicolumn{6}{l}{\textbf{Baselines}} \\
         & BM25     & 2 GB & -- &   .501 & .689 & .739 & .241 & .194 & .868 \\

         & BERT-Base$_\text{DOT}$ ANCE \cite{xiong2020approximate}     && -- &   .648 & -- & -- & -- & .330 & .959 \\
         & TCT-ColBERT \cite{lin2020distilling}   && -- &   .670 & -- & .720 & -- & .335 & .964 \\
         & RocketQA \cite{ding2020rocketqa}       && -- &   -- & -- & -- & -- & .370 & .979 \\

        \midrule
         \multicolumn{6}{l}{\textbf{Our Dense Retrieval Student Models}} \\
         \arrayrulecolor{lightgray}

         &\multirow{3}{*}{BERT-Base$_\text{DOT}$}   & \multirow{3}{*}{12.7 GB}& -- &  .593 & .757 & .664 & .347 & .294 & .913 \\
         &                                          && T1 &  .631 & .771 & .702 & .358 & .304 & .931 \\
         &                                          && T2 &  \textbf{.668} & \textbf{.826} & \textbf{.737} & \textbf{.371} & \textbf{.315} & \textbf{.947} \\

         \midrule 
         &\multirow{3}{*}{DistilBERT$_\text{DOT}$}  & \multirow{3}{*}{12.7 GB}& -- & .626 & .836 & .713 & .354 & .299 & .930 \\
         &                                          && T1 & .687 & .818 & .749 & .379 & .321 & .954 \\
         &                                          && T2 & \textbf{.697} & \textbf{.868} & \textbf{.769} & \textbf{.381} & \textbf{.323} & \textbf{.957} \\

        \arrayrulecolor{black}
        \bottomrule
    \end{tabular}
    \vspace{-0.3cm}
\end{table*}

To also compare the $\bertdot$ model in the full collection vector retrieval setting we set out to answer: 
\RQtwoB
The difference to previous results in Table \ref{tab:all_results} is that now we only use the score of a nearest neighbor search of all indexed passages, without re-ranking BM25. Because we no longer re-rank first-stage results, the pipeline overall becomes more efficient and less complex, however the chance of false positives becomes greater and less interpretable in a dense vector space retrieval. The $\colbert$ architecture also includes the possibility to conduct a dense retrieval, however at the expense of increasing the storage requirements of 2GB plain text to a 2TB index, which stopped us from conducting extensive experiments with $\colbert$.

We show nearest neighbor retrieval results of our $\bertdot$ models (using both BERT-Base and DistilBERT encoders) and baselines for dense retrieval in Table \ref{tab:dense_results}. Training with a teacher ensemble is again more effective than training with a single teacher, which is still more effective than training the $\bertdot$ alone without teachers. Interestingly, DistilBERT outperforms BERT-Base across the board with half the Transformer layers. As we let the models train as long as they improved the early stopping set, it suggests, for the retrieval task we may not need more model capacity, which is a sure bet to improve results on the $\bertcat$ architecture.

Our dense retrieval results are competitive with related methods, even though they specifically train for the dense retrieval task. Our approach, while not specific to dense retrieval training is competitive with the more costly and complex approaches ANCE and TCT-ColBERT. On MSMARCO DEV MRR@10 we are at a slight disadvantage, however we outperform the models that also published TREC-DL'19 results. RocketQA, the current state-of-the-art dense retrieval result on MSMARCO DEV requires a batch size of 4,000 and enormous computational resources, which are hardly comparable to our technique that only requires a batch size of 32 and can be trained on a single GPU.

\begin{figure}[t]
   \includegraphics[trim={0cm 0cm 0cm 0cm},width=0.37\textwidth]{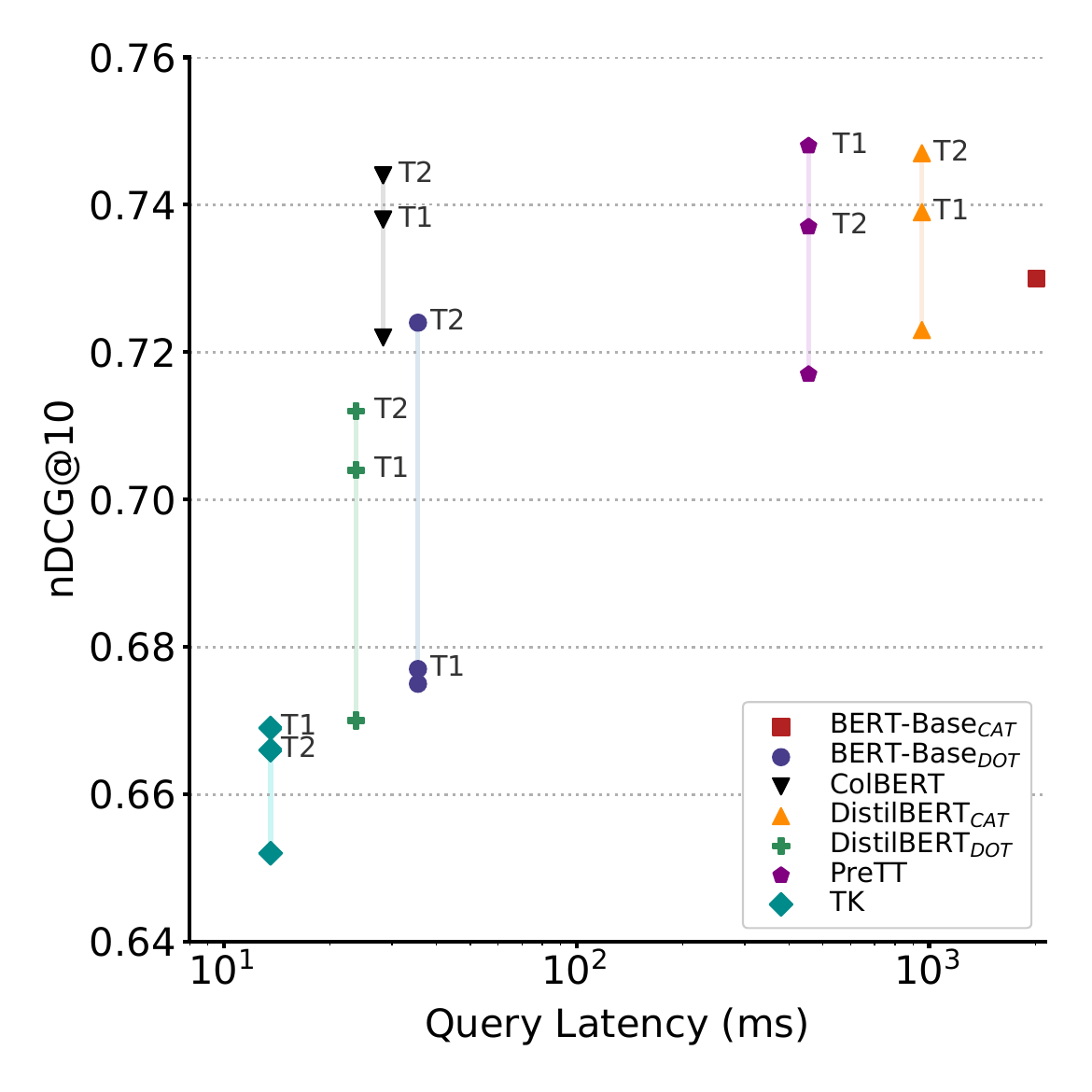}
    \centering
    \vspace{-0.5cm}
    \caption{Query latency vs. nDCG@10 on TREC'19 }
    \label{fig:ql-trec}
    \vspace{-0.1cm}
\end{figure}

\begin{figure}[t]
   \includegraphics[trim={0cm 0cm 0cm 0cm},width=0.37\textwidth]{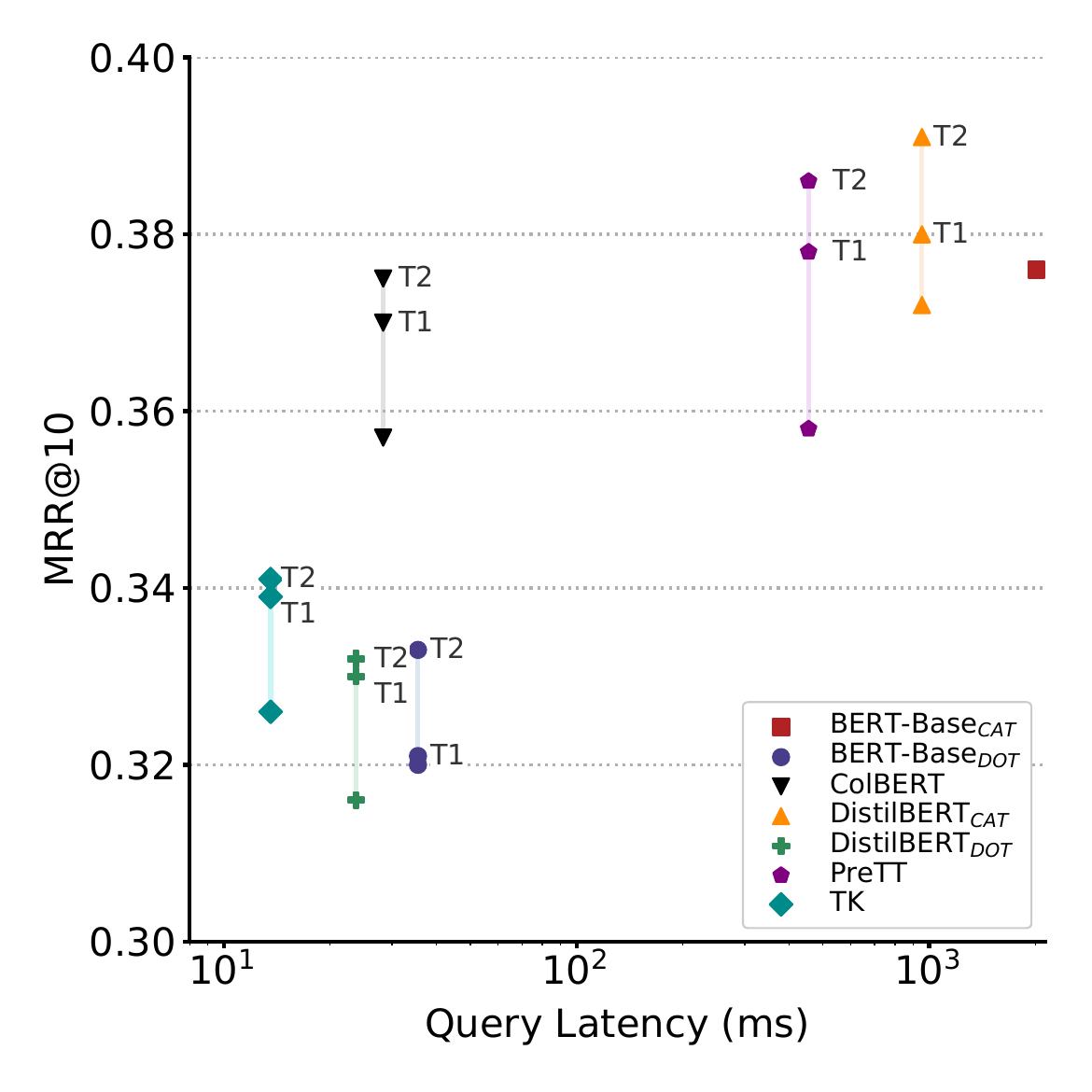}
    \centering
    \vspace{-0.5cm}
    \caption{Query latency vs. MRR@10 on MSMARCO DEV}
    \label{fig:ql-msmarco}
\end{figure}

\subsection{Closing the Efficiency-Effectiveness Gap}

We round off our results with a thorough look at the effects of knowledge distillation on the relation between effectiveness and efficiency in the re-ranking scenario. We measure the median query latency under the conditions that we have our cached document representation in memory, contextualize a single query, and computed the respective model's interaction pattern for 1 query and 1000 documents in a single batch on a TITAN RTX GPU with 24GB of memory. The large GPU memory allows us to also compute the same batch size for $\bertcat$, which for inference requires 16GB of total reserved GPU memory in the BERT-Base case. We measure the latency of the neural model in PyTorch inference mode (without accounting for pre-processing or disk access times, as those are highly dependent on the use of optimized inference libraries) to answer:

\RQthree

In \Cref{fig:ql-trec,fig:ql-msmarco}, we plot the median query latency on the log-scaled x-axis versus the effectiveness on the y-axis. The teacher trained models are indicated with \textit{T1} and \textit{T2}. The latency for different teachers does not change, as we do not change the architecture, only the weights of the models. The T1 teacher model $\bertcat$ is indicated with the red square. The TREC-DL'19 results in \Cref{fig:ql-trec} show how DistilBERT$_\text{CAT}$, PreTT, and ColBERT not only close the gap to BERT-Base$_\text{CAT}$, but improve on the single instance BERT-Base$_\text{CAT}$ results. The $\bertdot$ and TK models, while not reaching the effectiveness of the other models, are also improved over their baselines and are more efficient in terms of total runtime (TK) and index space ($\bertdot$). The MSMARCO DEV results in \Cref{fig:ql-msmarco} differ from \Cref{fig:ql-trec} in DistilBERT$_\text{CAT}$ and PreTT outperforming BERT-Base$_\text{CAT}$ as well as the evaluated $\bertdot$ variants under-performing overall in comparison to TK and ColBERT.

Even though in this work we measure the inference time on a GPU, we believe that the most efficient models --- namely TK, ColBERT, and $\bertdot$ --- allow for production CPU inference, assuming the document collection has been pre-computed on GPUs. Furthermore, in a cascading search pipeline, one can \textit{hide} most of the remaining computation complexity of the query contextualization during earlier stages.

\section{Teacher analysis}

Finally, we analyse the distribution of our teacher score margins, to validate the intuition of using a teacher ensemble and we look at per-query nDCG changes for two models between teacher-trained instances and the baseline. 

\subsection{Teacher Score Distribution Analysis}
\label{sec:teacher-score-analysis}

\begin{figure}[t]
   \includegraphics[trim={0cm 0cm 0cm 0cm},width=0.45\textwidth]{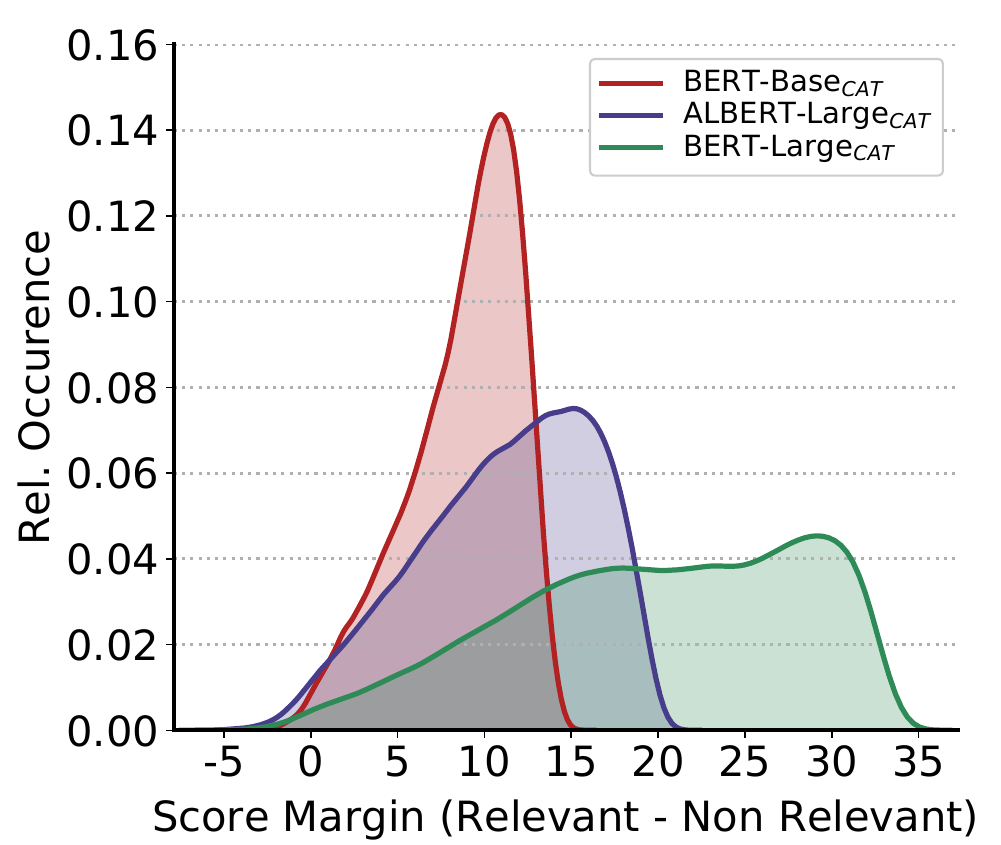}
    \centering
    \caption{Distribution of the margins between relevant and non-relevant documents of the three teacher models on MS MARCO-Passage training data}
    \label{fig:teacher_score_dist}
    \vspace{-0.2cm}
\end{figure}

To validate the use of an ensemble of teachers for RQ2, we analyze the output score margin distribution of our teacher models in Figure \ref{fig:teacher_score_dist}, to see if they bring diversity to the ensemble mix. This is the margin used in the Margin-MSE loss. We observe that the same $\bertcat$ architecture, differing only in the BERT language model used, shows three distinct score patterns. We view this as a good sign for the applicability of an ensemble of teachers, indicating that the different teachers have different viewpoints to offer. To ensemble our teacher models we computed a mean of their scores per example used for the knowledge distillation, to not introduce more complexity in the process.

An interesting quirk of our Margin-MSE definition is the possibility to reverse orderings if the margin between a pair is negative. In Figure \ref{fig:teacher_score_dist} we can see the reversal of the ordering of pairs in the distribution for the < 0 margin. It happens rarely and if a swap occurs the score difference is small. We investigated this issue by qualitatively analyzing a few dozen cases and found that the teacher models are most of the time correct in their determination to reverse or equalize the margin. Because it only affects a few percent of the training data we retained those samples as well to not change the training data.

\subsection{Per-Query Teacher Impact Analysis}
\label{sec:teacher-score-analysis}

\begin{figure}[t]
   \includegraphics[trim={0.4cm 0cm 0.3cm 0cm},width=0.47\textwidth]{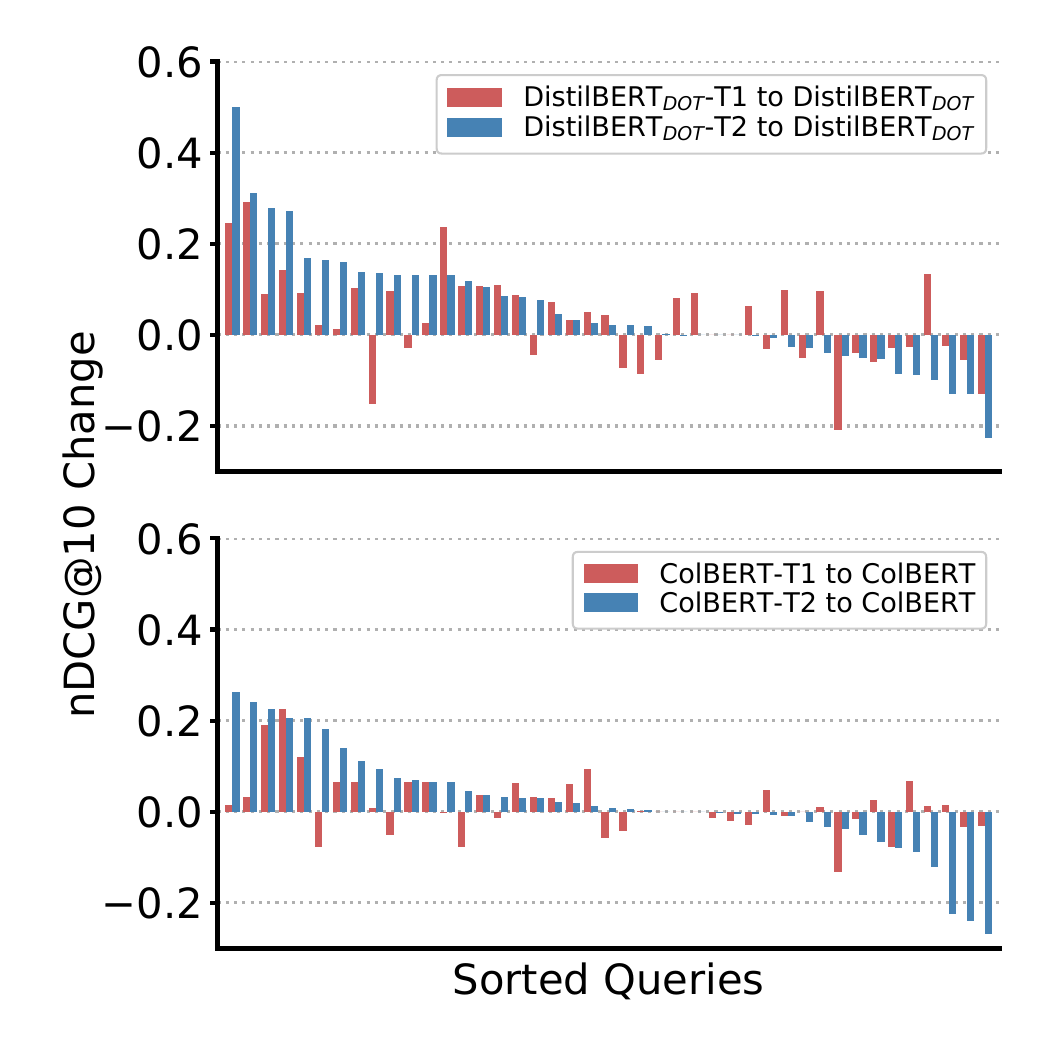}
    \centering
    \vspace{-0.6cm}
    \caption{A detailed comparison between T1 and T2 training ndcg@10 changes per query of the TREC-DL'19 query set}
    \label{fig:per-query-changes}
    \vspace{-0.2cm}
\end{figure}

In addition to the aggregated results presented in Table \ref{tab:all_results}, we now take a closer look at the impact of T1 and T2 teachers in a per-query analysis for ColBERT and DistilBERT$_\text{DOT}$ in Figure \ref{fig:per-query-changes}. We plot the differences in nDCG@10 per query on the TREC-DL'19 set between the original training results and the T1 and T2 training respectively. A positive change means the T1/T2 trained model does better on this particular query. We sorted the queries by the T2 changes for both plots, and plotted the corresponding query results for T1 at the same position. 
Overall, the T1 \& T2 training for both models roughly improves 60 \% of queries and decreases results on 33 \% with the rest of queries unchanged. Interestingly, the average change in each direction between the T1 and T2 training shows that T2 results become more extreme, as they improve more on average (DistilBERT$_\text{DOT}$ from T1 $+10$\% to T2 $+13$\%; ColBERT from T1 $+6$\% to T2 $+9$\%), but also decrease stronger on average (DistilBERT$_\text{DOT}$ from T1 $-6.8$\% to T2 $-7.2$\%; ColBERT from T1 $-4.3$\% to T2 $-7.8$\%). As we saw in Table \ref{tab:all_results} the aggregated results, still put T2 in front of T1 overall. However, we caution, that these stronger decreases show a small limitation of our knowledge distillation approach.












\section{Related Work}

\paragraph{\textbf{Efficient relevance models}}
Recent studies have investigated different approaches for improving the efficiency of relevance models. \citet{ji2019efficient} demonstrate that approximations of interaction-based neural ranking algorithms using kernels with locality-sensitive hashing accelerate the query-document interaction computation. 
In order to reduce the query processing latency, \citet{mackenzie2020efficiency} propose a static index pruning method when augmenting the inverted index with precomputed re-weighted terms \cite{dai2020context}.
Several approaches aim to improve the efficiency of transformer models with windowed self-attention \cite{Hofstaetter2020_sigir}, using locality-sensitive hashing \cite{kitaev2020reformer}, replacing the self-attention with a local windowed and global attention \cite{beltagy2020longformer} or by combining an efficient transformer-kernel model with a conformer layer \cite{mitra2020conformer}.

\paragraph{\textbf{Adapted training procedures}}
In order to tackle the challenge of a small annotated training set, \citet{dehghani2017learning} propose weak supervision controlled by full supervision to train a confident model. 
Subsequently they demonstrate the success of a semi-supervised student-teacher approach for an information retrieval task using weakly labelled data where the teacher has access to the high quality labels \cite{dehghani2017fidelity}. 
Examining different weak supervision sources, \citet{macaveney2020headings} show the beneficial use of headline - content pairs as pseudo-relevance judgements for weak supervision.
Considering the success of weak supervision strategies for IR, \citet{khattab2020colbert} train ColBERT \cite{khattab2020colbert} for OpenQA with guided supervision by iteratively using ColBERT to extract positive and negative samples as training data.
Similarly \citet{xiong2020approximate} construct negative samples from the approximate nearest neighbours to the positive sample during training and apply this adapted training procedure for dense retrieval training. 
\citet{cohen2019} demonstrate that the sampling policy for negative samples plays an important role in the stability of the training and the overall performance with respect to IR metrics. 
\citet{macavaney2020training} adapt the training procedure for answer ranking by reordering the training samples and shifting samples to the beginning which are estimated to be easy.

\paragraph{\textbf{Knowledge distillation}}

Large pretrained language models advanced the state-of-the-art in natural language processing and information retrieval, but the performance gains come with high computational cost. There are numerous advances in distilling these models to smaller models aiming for little effectiveness loss.  

Creating smaller variants of the general-purpose BERT mode, \citet{jiao2019tinybert} distill TinyBert and \citet{sanh2019distilbert} create DistilBERT and demonstrate how to distill BERT while maintaining the models' accuracy for a variety of natural language understanding tasks.

In the IR setting, \citet{tang2018ranking} distill sequential recommendation models for recommender systems with one teacher model. \citet{vakili2020distilling} study the impact of knowledge distillation on BERT-based retrieval chatbots.
\citet{gao2020understanding} and \citet{chen2020simplified} distilled different sizes of the same $\bertcat$ architecture and the TinyBert library \cite{jiao2019tinybert}.
As part of the PARADE document ranking model \citet{li2020parade} showed a similar $\bertcat$ to $\bertcat$ same-architecture knowledge distillation for different layer and dimension hyperparameters.
A shortcoming of these distillation approaches is that they are only applicable to the same architecture which restricts the retrieval model to full online inference of the $\bertcat$ model.
\citet{lu2020twinbert} utilized knowledge distillation from $\bertcat$ to $\bertdot$ in the setting of keyword matching to select ads for sponsored search. They first showed, that a knowledge transfer from $\bertcat$ to $\bertdot$ is possible, albeit in a more restricted setting of keyword list matching in comparison to our fulltext ranking setting.

\section{Conclusion}

We proposed to use cross-architecture knowledge distillation to improve the effectiveness of query latency efficient neural passage ranking models taught by the state-of-the-art full interaction $\bertcat$ model. Following our observation that different architectures converge to different scoring ranges, we proposed to optimize not the raw scores, but rather the margin between a pair of relevant and non-relevant passages with a Margin-MSE loss. We showed that this method outperforms a simple pointwise MSE loss. Furthermore, we compared the performance of a single teacher model with an ensemble of large $\bertcat$ models and find that in most cases using an ensemble of teachers is beneficial in the passage retrieval task.
Trained with a teacher ensemble, single instances of efficient models even outperform their single instance teacher models with much more parameters and interaction capacity. We observed a drastic shift in the effectiveness-efficiency trade-off of our evaluated models towards more effectiveness for efficient models. In addition to re-ranking models, we show our general distillation method to produce competitive effectiveness compared to specialized training techniques for the dual-encoder $\bertdot$ model in the nearest neighbor retrieval setting. We published our teacher training files, so the community can use them without significant changes to their setups. For future work we plan to combine our knowledge distillation approach with other neural ranking training adaptations, such as curriculum learning or dynamic index sampling for end-to-end neural retrieval. 

\balance
\bibliographystyle{ACM-Reference-Format}
\bibliography{references}

\end{document}